\documentclass[a4paper]{jpconf}
\usepackage{graphicx,
            amsmath,
            amssymb,
            hyperref}
\begin{document}

\title{An introduction to computational complexity and statistical learning theory applied to nuclear models}

\author{Andrea Idini}

\address{Division of Mathematical Physics, Physics dept., LTH, Lund University, S-22100 Lund, Sweden}

\ead{andrea.idini@matfys.lth.se}

\begin{abstract}
The fact that we can build models from data, and therefore refine our models with more data from experiments, is usually given for granted in scientific inquiry. However, how much information can we extract, and how precise can we expect our learned model to be, if we have only a finite amount of data at our disposal? 
Nuclear physics demands an high degree of precision from models that are inferred from the limited number of nuclei that can be possibly made in the laboratories.

In manuscript I will introduce some concepts of computational science, such as statistical theory of learning and Hamiltonian complexity, and use them to contextualise the results concerning the amount of data necessary to extrapolate a mass model to a given precision.
\end{abstract}

\section{Introduction}

After more than 80 years from the first models describing the nuclear binding energies, deviations from experiments of nuclear mass models ($\approx$ MeV) are orders of magnitude larger than experimental errors in the measurements of masses ($\lesssim$ keV). It seems extremely difficult to substantially improve over effective phenomenological models such as Duflo--Zucker \cite{Duflo:95} or finite range droplet model \cite{Moller:12}, despite the several efforts driven by the need to investigate several phenomena related to nuclei not yet measured, such as those that occur in astrophysical environments.

To the this present day, attempts to devise a precise model of the ground state properties of nuclei continue to build upon previous successes. To name a few strategies adopted: expanding with additional corrections and procedures the aforementioned mass models \cite{Liu:11, Ye:21, Verriere:21}; considering a formally consistent and, in principle, systematically improvable energy density functional in the Skyrme \cite{ryssens2022skyrme}, or Gogny \cite{Goriely:09, goriely2016gogny}, or relativistic \cite{pena2016relativistic}, or new and higher order \cite{Davesne:13, Bennaceur:17} functional forms; using a statistical and machine learning approaches to redefine nuclear models in terms of correlations \cite{Niu:18,Shelley:21, Liu:21} (cf. the recent review \cite{Boehnlein:22}). In these remarkable advances the improvement in precision and reliability of extrapolation has been steady, but arguably insufficient for a radical change in our understanding of the nuclear system as of yet. In fact, the two particular models  \cite{Duflo:95, Moller:12} are still used in state--of--the--art calculations, e.g. of astrophysical r--process nucleosynthesis (cf. the recent review \cite{Cowan:21} and refs. therein).

I previously analysed mass models and energy density functionals with the tools provided by the statistical learning theory and hence argued  that this difficulty is not the shortcoming of any specific model or statistical approach, but linked to the fundamental nature of model building as based on data extrapolation \cite{Idini:20}. The degree of precision required by a sizeable improvement over current models cannot be reached considering models trained on the limited the ground state data of nuclei.

In this proceeding, I will take the opportunity to first put this result in the larger context of the application of theoretical information science to physical models. Then, I will introduce statistical learning and probably approximately correct learning with an example, and finally summarize the main results of \cite{Idini:20} related to nuclear mass models.

\section{What a model needs}
% What is a model?
A model is a mathematical construction that considers available information and processes it. The information a model is built upon is coming from data, their context, and previous models and theories that the model builder assumes can be generalized to the problem under consideration.  Scientifically, the objective of a model is to explain and predict new results. We consider this endeavour an essential efficient way of understanding physical phenomena. 
It is useful to the discussion in this manuscript to consider that the objective of a model is to yield predictions and interpretations, not a beautiful theory. In the words of George Box: "{\it all models are wrong; the practical question is how wrong do they have to be to not be useful}."

Concerning nuclear physics, quantitative models are used describe and predict the properties of the atomic nuclei, comparing the results to the available data (postdiction) and with data that will be measured (prediction). It is useful to notice that models of nuclei are not exclusively made from data concerning nuclei, but also on the knowledge and context concerning the data itself, quantum mechanics, and fundamental interactions. 
% Our confidence in a model doesn't restrict to atomic nuclei in particular but extends e.g. to many--body quantum systems in general. 
Nuclear physics and models are eventually analysed through the lens of statistics to evaluate the expected precision and agreement with experimental results in a rigorous way \cite{Dobaczewski:14}.

% What makes a method being able to have results?

The mathematical construction of the model will need two things to process information and be useful: data to calibrate the model and computation to yield results. It is possible to evaluate the limits imposed to specific models with respect to these two needs, at least under certain conditions. What is the necessary amount of operations needed to apply the model to compute a given case, and how would that scale to other cases? What is the necessary amount of data to calibrate a model to a given precision, and eventually falsify it? Since all models are wrong, if a model cannot be constrained to the required precision or cannot compute the required cases, it is both wrong and useless. Let's consider those limits that can help identify useful models.

\subsection{Hamiltonian complexity}

% What does it mean that something is computationally feasible?
The application of a model needs computation to yields results. This computation can have varying degree of complexity, from a low rank polynomial like the semi--empirical mass formula to the most complicated no-core shell model requiring the largest supercomputer. {\it Computational complexity} is a familiar tool in computer science that estimates how the resources required by algorithm, either memory or number of operations, scale. The computational scaling of an algorithm can be assessed at different levels of abstraction, from the profiling and scaling of a specific implementation of an algorithm to the general asymptotical behaviour of a whole class of solutions to problems. For example, different algorithms to sort an array can have different scaling with respect to the number $n$ of elements in an array, famously $O(n\log(n))$ and $O(n^2)$ for quicksort and bubble sort respectively. However, if we are only interested in knowing whether the computational capacity will grow faster than the difficulty of a problem, and therefore if a problem can be {\it effectively} solvable for every chosen dimension $n$, we might consider only the {\it complexity class} of the sorting problem itself which is polynomial or {\bf P}.

Finding the complexity class pertains to certain problems rather than specific algorithmic implementation. In the same way it can be asked what is the complexity class related to a sorting problem without specifying or attempting a solution, it can be asked what is the complexity class of finding eigenstates of Hermitian operators in an Hilbert space. Using techniques from computer science and quantum computing, the field of Hamiltonian complexity investigates what complexity is related to specific many--body solutions of Hamiltonians (\cite{Osborne:12, McArdle:20} and refs. therein). 

Very notably for the efforts of nuclear physics, it was analysed the complexity of finding the solution to a local Hamiltonian with $k$--body interactions.  Common misconceptions attribute this problem either an exponential {\bf EXP} or a polynomial {\bf P} difficulty, coming either from considering the exponential number of possible Slater determinants or the diagonalization of the Hamiltonian matrix respectively. This problem is in fact demonstrated to be somewhat in the middle: having an oracle that provides an ansatz, a quantum computer can polynomially verify whether this ansatz is eigenstate of the Hamiltonian or no. This is denoted as quantum Merlin Arthur (due to the relation between an oracle called Merlin and a capable verifier Arthur), and indicated as {\bf QMA}. 
In \cite{Kempe:06} was demonstrated that the 2--body Hamiltonian is {\bf QMA}--complete. {\it Complete} means that any other {\bf QMA} problem, including a general $k$--body Hamiltonian, can be reduced to that.
Incidentally, it also means that quantum computers will have difficulty providing solutions to Hamiltonian problems in the same way classical computers have difficulty with non--deterministic polynomial {\bf NP}--complete problems such as the finding the relaxation to  minimum energy of a frustrated spin glass or protein (cf. Fig. \ref{fig:scheme}).

While Hamiltonian complexity indicates how difficult it is to find truly first principle solutions to a general Hamiltonian in a many--body context, a first principle to nuclear physics approach could still rely on the density functional theory. The Hohenberg--Kohn theorem guarantees that for every $k$--body Hamiltonian it exists a functional of the density enclosing all the ground state observables. 
Unfortunately, also density functional theory (DFT) was demonstrated to reside in {\bf QMA} complexity \cite{Schuch:09}.
While the calculation of the ground state density and observables using a given functional is only {\bf P}, the derivation and identification of the universal density functional belongs to {\bf QMA}, and is demonstrated in {\bf QMA}--complete for many specific cases \cite{Whitfield:13, OGorman:21}. Therefore, the promising route of identifying a universal nuclear density functional is intrinsically more computationally challenging than first expected.

\subsection{Data complexity}
% What is the data content required for a certain model

A model requires also to be calibrated with available data in order to be useful in explaining or predicting other data. It is well known that the more {\it complicated} a model is, the more data it requires to be constrained. 
% I have been told that a polynomial fit with $n$ parameters requires at least $n$ non--degenerate data points. However, in machine learning it is often the case that the model has more parameters than the data points available reaching excellent results. How is it possible?
Statistical learning theory formally studies this requirement of models, noting that the relation between model complexity and the data availability translates in an upper limit to the precision of generalization of a model to unknown data \cite{Vapnik:92,Vapnik:99}.

\begin{figure}[h!]
\centering
\includegraphics[height=100px]{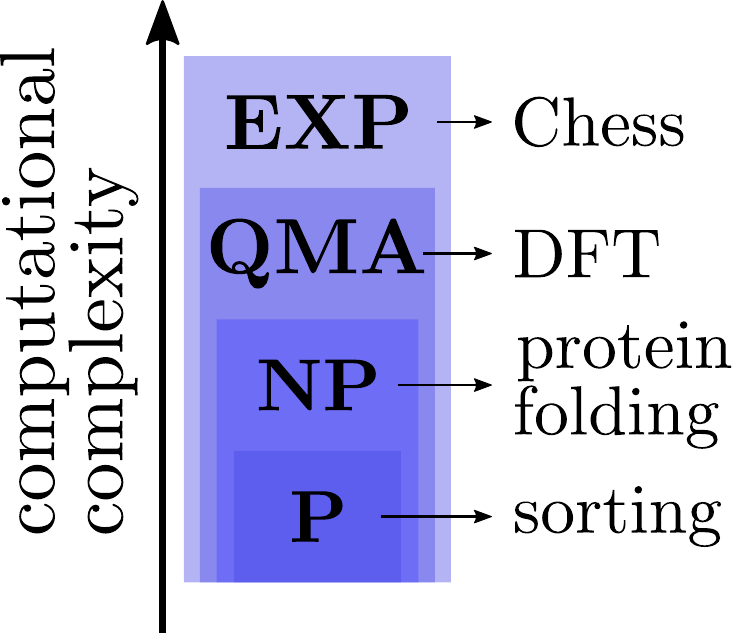} \qquad \qquad \includegraphics[height=100px]{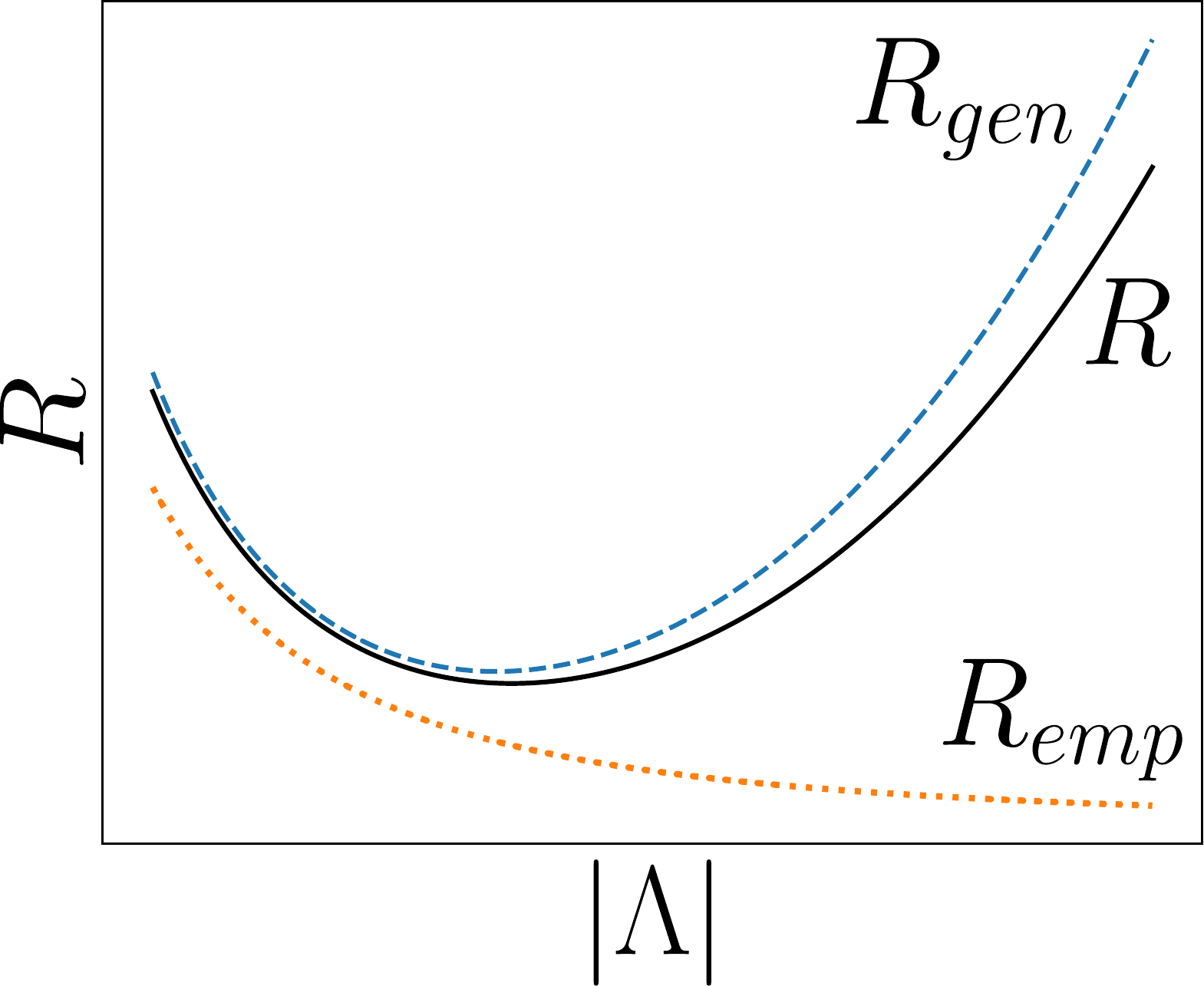} 
\caption{(left panel) A schematic representation of the different sets of complexity classes {\bf P $\subseteq$ NP $\subseteq$ QMA $\subseteq$ EXP} and relevant examples of members of each class. (right panel) schematic representation of the relation between the complexity of a model measured with the model dimension $|\Lambda|$ and the generalization, empirical, and total risks.}
\label{fig:scheme}
\end{figure}

Let's consider a theory corresponding to a class of models $\Lambda$ that aims to reproduce and predict data, for example the Skyrme functional generator form. Different instances of models $\alpha$ within this class will have a different set of parameters, as for one of the many variations and parametrizations. Generally, a model builder has the objective of finding out the "best" model within the class by minimizing a cost or risk functional $R(\alpha)$. However, when working with a finite data set what is actually minimized is the empirical risk $R_{emp(l)}(\alpha)$, the risk evaluated over a finite number of training data $l$. The error on other measurements not used for the training of the model, is denoted as generalization error $R_{gen}(\alpha)$. By definition, we cannot really fully know $R_{gen}(\alpha)$ and a model has only a probability to appropriately generalize to data not measured, yielding $R(\alpha) \approx R_{emp(l)}(\alpha) \approx R_{gen}(\alpha)$ (cf. Fig. \ref{fig:scheme}). 

Physical models are usually trained not only to predict new measurements, but importantly to the much more difficult task of predicting new measurements outside of the conditions presently available. This is particularly important in the case of nuclear mass models, where need for interpolation is rare or non--existent and the objective is to predict the binding energy for nuclei with more neutrons and/or protons than any previously measured. Furthermore, models of the nuclear masses have the quite unusual feature that not only the available data is limited, but also the total number of possible data is finite.

The complexity of the model is crucial to evaluate its reliability in the case of a finite amount of data. Conventionally, it is clear that a “simple” model has a better chance than a “complicated” to generalize well. Philosophically, a simple model is preferable than a complicated one, following the Occam's razor principle. Statistically this feature is understood as the complicated model is prone to overfitting. Physicists often consider a simple model as more desirable and deserving attention, especially if it is also rooted in theories outside of the present limited scope. For example, models that effectively take into account several observables and properties of nuclei such as \cite{Duflo:95, Moller:12, Goriely:09} are widely and rightfully considered more reliable than purely data--driven approaches. All of these perspectives can be incorporated and analysed within the framework of statistical learning theory. I proposed in \cite{Idini:20} to use statistical learning theory to investigate the generalization error of mass models and nuclear density functionals and in the following will provide more examples.

\section{Statistical learning bounds}

% What is the relation between the number of data points, and the complexity?
Wihtout loss of generality, I will consider the example case of extrapolating of a mass model to unknown nuclei. In this case, models (also called hypotheses) $\alpha$ belonging to the model class $\Lambda$ will be evaluated, e.g. specific Skyrme parametrizations belonging to the class of all Skrme density functionals. Let me further simplify the problem by just asking if a measured binding energy is within a reasonable error expected from the model or not, instead of evaluating the absolute value of the error. In other words, associate a certain probability for the new data point to disagree with the model within this error $\epsilon$, e.g. $\epsilon \approx 0.05$ assuming a normal distribution the errors and 2$\sigma$ tolerance. The following derivation will be the same in the case of more complicated assumptions.

If the model is a "good" model, so that $R(\alpha') \approx R_{emp(l)}(\alpha')$, an experiment will yield a measurement within error with the attributed probability $1 - \epsilon$. In the case of a "bad" model, the probability that a bad hypothesis $\alpha_{bad}$ calculated for $x_1$ will agree with the corresponding data point $y_1$ is less than expected, $P[\alpha_{bad}(x_1) = y_1] < 1-\epsilon$. For $m$ independent data points, this implies $P[\alpha_{bad}(x_1) = y_1, \cdots, \alpha_{bad}(x_m) = y_m] < (1-\epsilon)^m < e^{-m\epsilon}$. This is known as the {\it Hoeffding's inequality}. If we are considering all the models $\alpha$ within the class $\Lambda$, there is a probability $\delta$ that an $\alpha$ will generalize badly, that it will agree with training and disagree with an external additional data point, $\delta \equiv P[\alpha(x_1) = y_1, ..., \alpha(x_m) = y_m, \alpha(\tilde x) \neq \tilde y]$. This probability will depend on the {\it degrees of freedom} that $\alpha$ has to adapt to data $x_1, \cdots, x_m$, while still missing $\tilde x$. From the previous relations and, it is possible to assess the probability of a "good" $\alpha$ that generalizes well instead,
\begin{align}
\delta \leq \sum_{\alpha \in \Lambda} 
P[\alpha(x_1) = y_1, \cdots, \alpha(x_m) = y_m, \alpha(\tilde x) = \tilde y] \leq |\Lambda| e^{-m\epsilon}\;,
\label{eq:Hoeffding}
\end{align}
with $|\Lambda|$ comes from the sum over the possible models and represents the dimension of the hypothesis space, which we will define more precisely later. The Hoeffding's inequality therefore can be rewritten to provide a relation between the number of data points $m$ needed to possibly reach a generalization error $\epsilon$ with probability $1-\delta$. This is obtained considering the probability $\delta$ that a model hypothesis $\alpha$ within the space $\Lambda$ fits with $l$ training data points and doesn't fit with a point not belonging to the training set. This is called probably approximately correct (PAC) learning.

% ----------------- %

% How is the complexity of a model measured?

There are several ways to quantitatively estimate the dimension of a model space $|\Lambda|$. The number of parameters is often used, and it is a good indicator in case of polynomials. However, it is known that in general some models with many parameters can be well constrained, while one single free parameter can be extremely adjustable. For a better estimate, I used in \cite{Idini:20} the Vapnik--Chervonenkis (VC) dimension \cite{Vapnik:92} that considers the flexibility of a model to adapt in order to accommodate different data results. For a review of different possible method to evaluate the complexity of a statistical learning model cf. \cite{Abbas:21}.

Therefore, in a PAC learning setting the number of data $m$ to reach a certain precision $\epsilon$ is given by Eq. (\ref{eq:Hoeffding}) to be at least,
\begin{equation}
m \geq \frac{1}{\epsilon}\left[ \log(|\Lambda|) + \log \left( \frac{1}{\delta} \right) \right],
\label{eq:m}
\end{equation}
with the size of the model class under consideration $|\Lambda|$, eventually taken as the VC--dimension. These are denoted as PAC learning bounds. 

% Why a general model is better
% Within this context it is also easy to see why a more general model is interesting and physics has worked so well
% Within this context a mass model that not only satisfies the good statistical indication of being well generalizable, but is based on solid physical principles and has insight inside other observables, is much more reliable than a purely statistical model. It is an indirect evaluation and expansion of the empirical risk, hence improving $m$ while keeping $\log(|\Lambda|)$ contained.

In \cite{Idini:20} I took these concepts and derived the PAC learning error bounds for Weiszacker mass formula, as an example of a simple phenomenological model, and few instances of nuclear density functionals. I obtained that the PAC generalization error bounds are $\gtrsim 1$ MeV, which is similar to the present root mean square deviation of the models. Therefore, this result provides an indication of the difficulty to improve upon these models based on statistical approaches alone, due to the data complexity of these models requiring more data points than those available.

\section{Discussion}

% Information content and context of scientific theories

The concepts of theoretical computer science described in this proceeding are widely used fundamental tools to think about computational models. Of course, the application of these general concepts to nuclear physics and the specific problem of nuclear mass models has to take into account important limitations of this analysis. 
The computational complexity identifies only the general scaling behaviour, which is crucial in the applications of an algorithm to asymptotically larger cases. However, the nuclear system is finite. Despite the fact that the solution to a general Hamiltonian or the derivation of the corresponding universal density functional belong to {\bf QMA} complexity class, making them substantially "unsolvable", it is possible that a specific implementation on a given hardware could provide solutions all possible cases for most observables of interest. 
Similarly, the PAC learning bounds are derived considering uncorrelated data points and without taking into account that theories of atomic nuclei have components that are more general and thoroughly tested than the limited nuclear mass model. Since it holds information, this scientific context can be used by the model builder, but it is difficult to rigorously quantify within the bounds of statistical learning theory.

Within this context, density functional theory presents an interesting opportunity to evaluate both the computational complexity and the data complexity in a robust setting. Other specific many--body methods (e.g. configuration interaction, coupled cluster, or generator coordinate method) could be also evaluated with respect to their computational and data complexity in future work. The relation between data complexity and computational complexity is still unexplored in this physics context, and could provide further insight into the workings and expected performance of models.

In conclusion, theoretical work on Hamiltonian complexity and statistical learning theory can inform on the most promising route for constructing models of nuclei and navigate the tradeoffs necessary in their practical implementations. 
One might notice that some of the limitations of Hamiltonian and complexity analysis are less severe, hence the conclusions of the theories more robust, in the case of models that aim to be "first principle" or "consistent". For example, computing solutions to Hamiltonians and statistically rigorous descriptions might have a tendency to keep the modeller outside of the model, to be as objective as possible.
These approaches are certainly fundamentally important for the development of nuclear physics. However, I hope that with this contribution I could provide some formal arguments in favour of complementary approaches, and frame the space for the role of the modeller in the construction of highly performing models.

\section*{References}

\bibliographystyle{iopart-num}
\bibliography{./nuclear, ../../neural}
\end{document}